\documentclass[conference]{IEEEtran}

\usepackage[sort,compress]{cite}
\usepackage{amsmath,amssymb,amsfonts}
\usepackage{algorithmic}
\usepackage{graphicx}
\usepackage{textcomp}
\usepackage{xcolor}
\usepackage{xspace}
\usepackage{physics}
\usepackage[procnumbered,linesnumbered,ruled]{algorithm2e}
\usepackage{subcaption}
\usepackage{xfrac}
\usepackage{amssymb}

\newcommand{\povm}{\texttt{QSD-Two}\xspace}
\newcommand{\povmone}{\texttt{QSD-One}\xspace}
\newcommand{\pqcone}{\texttt{PQC-One}\xspace}
\newcommand{\pqctwo}{\texttt{PQC-Two}\xspace}
\newcommand{\eat}[1]{}

\ifx\noeditingmarks\undefined
    
\else
    
\fi

\newcommand{\mpfont}{\scriptsize}

\ifx\noeditingmarks\undefined
    \newcommand{\MPworker}[2]{{\color{#1}\vrule\vrule}{\marginpar{\color{#1}\mpfont #2}}}
\else
    \newcommand{\MPworker}[2]{}
\fi

\newcommand{\lacc}{\mathrm{CC_{acc}}}
\newcommand{\lerr}{\mathrm{L_{err}}}

\newcommand{\para}[1]{\smallskip \noindent\textbf{#1}}
\newcommand{\softpara}[1]{\smallskip \noindent \underline{#1}}

\usepackage{url}

\begin{document}

\title{Quantum Sensor Network Algorithms for Transmitter Localization}

\author{\IEEEauthorblockN{Caitao Zhan}
\IEEEauthorblockA{\textit{Department of Computer Science} \\
\textit{Stony Brook University, USA}\\
}
\and
\IEEEauthorblockN{Himanshu Gupta}
\IEEEauthorblockA{\textit{Department of Computer Science} \\
\textit{Stony Brook University, USA}\\
}
}

\maketitle

\begin{abstract}
A quantum sensor (QS) is able to measure various physical phenomena with extreme sensitivity.
QSs have been used in several applications such as atomic interferometers, but few applications of a quantum sensor network (QSN) have been proposed or developed.
We look at a natural application of QSN---localization of an event (in particular, of a wireless signal transmitter).
In this paper, we develop effective quantum-based techniques for the localization of a
transmitter using a QSN.
Our approaches pose the localization problem as a well-studied quantum state discrimination (QSD) problem and address the challenges in its application to the localization problem. 
In particular, a quantum state discrimination solution can suffer from a high probability of 
error, especially when the number of states (i.e., the number of potential transmitter locations in our case) can be high. 
We address this challenge by developing a two-level localization approach, which localizes the transmitter at a coarser granularity in the first level, and then, in a finer granularity in the second level. 
We address the additional challenge of the impracticality of general measurements by 
developing new schemes that replace the QSD's measurement operator with a trained parameterized hybrid quantum-classical circuit.
Our evaluation results using a custom-built simulator show that our best scheme is able to 
achieve meter-level (1-5m) localization accuracy; 
in the case of discrete locations, 
it achieves near-perfect (99-100\%) classification accuracy. 


\end{abstract}



\section{\bf Introduction}


Quantum sensors, being strongly sensitive to external disturbances, are able to measure various physical phenomena with extreme sensitivity.
These quantum sensors interact with the environment and have the environment phenomenon or parameters encoded in their state~\cite{RevModPhys.quantumsensing}.
A group of distributed quantum sensors, if prepared in an appropriate entangled state, can further enhance the estimation of a single continuous parameter, improving the standard deviation of measurement by a factor of $1/\sqrt{m}$ for $m$ sensors (Heisenberg limit)~\cite{Giovannetti_2011}.
\eat{Recently, experimental physicists successfully demonstrated a reconfigurable distributed radio-frequency photonic sensor network~\cite{PRL20-qsn,arizona21-thesis} that utilizes squeezed quantum state and entanglement to enhance sensing of radio signals.}

Recently, many protocols have been developed for the estimation of a single 
parameter or multiple independent parameters~\cite{Giovannetti_2011,Proctor_2018} using one or multiple (possibly, entangled) sensors. 
But, the use of a distributed set of quantum sensors working collaboratively 
to estimate more complex physical/environmental phenomena, as in many classical
sensor network applications~\cite{tsn17-water, sensys10-health,mobicom03-sensor}, 
has not been explored much.
In this paper, we explore a potential quantum sensor network application--- localization of events.
In particular, we develop effective techniques 
for radio frequency (RF) transmitter localization and thus demonstrate the promise of QSNs in the accurate localization of events. Our motivation for choosing 
RF transmitter localization as the event localization 
application is driven by the significance of transmitter localization
in  wireless/mobile applications and recent advances in quantum sensor
technologies for RF signal detection (see \S\ref{sec:problem}).

\begin{figure*}[t]
    \centering
    \includegraphics[width=0.72\textwidth]{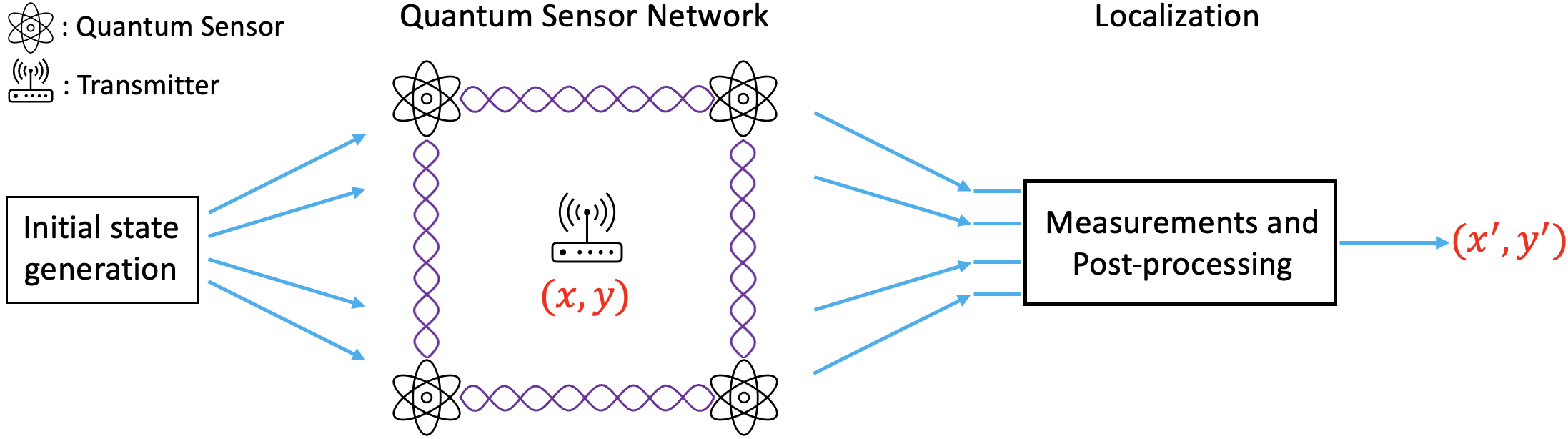}
    \caption{Overall architecture of using a QSN to localize a transmitter. 
    }
    \vspace{-0.1in}
    \label{fig:quantumoverall}
\end{figure*}

\para{Transmitter Localization using QSNs.}
Our approach to
transmitter localization using QSNs essentially involves posing
the localization problem as a quantum state discrimination
(QSD) problem~\cite{bergou-review-2007} which is to identify the specific state a given
quantum state is in (from a given set of states in which the
system can be) by performing quantum measurements on the
given quantum system. The overall architecture is illustrated in
Fig.~\ref{fig:quantumoverall}. 
First, a probe state is generated and distributed to the
QSN. Then, once the quantum sensors have been impacted
(i.e., the overall quantum state changed) due to the transmission
from the transmitter’s signal, an appropriate quantum measurement 
is made on the quantum state of the network.
The outcome of the measurement determines the quantum
state, and thus, the location of the transmitter. 
However, the
above process can be erroneous, as solving 
the QSD problem even optimally
can incur a certain probability of (classification/discrimination)  error.
This paper’s goal is to develop an approach with a 
minimal localization error. 
In that context, our
developed schemes in this paper are based on two ideas that extend
the above basic QSD-based approach: 
\begin{enumerate}
    \item We use a two-level approach
that localizes the transmitter in two stages: first, at a coarse level
using a set of sensors over the entire area, and then, at a fine level
within the ``block'' determined by the first level. 
\item In addition, we circumvent the challenge of implementing a general
measurement operation, by instead using a trained parameterized 
hybrid quantum-classical circuit that essentially implements the 
measurement operation and predicts the transmitter location from
quantum sensor data.
\end{enumerate}
Our evaluation results show that our best scheme (which combines the above two ideas) is able to achieve meter-level (1-5m) localization accuracy; in case of discrete locations, it achieves near-perfect (99-100\%) classification accuracy. 

\para{Contributions.} 
In the above context, we make the following contributions. 
\begin{enumerate}
    \item We model the transmitter localization problem as a well-studied quantum state discrimination (QSD) problem, which allows us to develop viable transmitter localization schemes using quantum sensors. 
 
    \item 
    We design two high-level schemes to localize a transmitter in a given area deployed with a quantum sensor network.
     The first scheme is based on solving an appropriate quantum state discrimination problem using a global measurement, while the second scheme uses a trained hybrid quantum-classical circuit to process the quantum sensor data. Within the above high-level schemes, we also introduce a two-level localization scheme to improve the performance of the basic one-level schemes.
  
    \item 
    To evaluate our schemes, we model how a quantum sensor's state evolves due to RF signals from a transmitter at a certain distance. Using this model, we 
     evaluate our localization schemes and demonstrate their effectiveness in our custom-built simulator.
\end{enumerate}


\para{Paper Organization.} The paper is organized as follows. 
In \S\ref{sec:problem}, we present our quantum sensor model, formally define the transmitter localization problem and discuss related work.
In the following two sections, we describe our two classes of algorithm: quantum-state-discrimination (QSD) based scheme, and parameterized-quantum-circuit (PQC) based scheme.
We discuss our evaluation results in \S\ref{sec:eval}, and give concluding remarks
in \S\ref{sec:conclusion}.

\section{\bf Sensor Model, Problem, and Related Work}
\label{sec:problem}

In this section, we start with motivating our choice of RF transmitter localization as an application for QSN, and then model the impact of an RF received signal on the quantum state of a quantum sensor. We then formulate the quantum localization problem and discuss related work.

\para{Motivation for Transmitter Localization.}
Accurate detection and localization of a wireless transmitter 
(typically, using a radio-frequency (RF) wireless signal) is important in a 
variety of wireless and/or mobile
applications, e.g.,
as an intruder detection in shared spectrum systems~\cite{infocom18-spectrum}, localization of devices/users in
indoor settings (e.g., supermarkets, museums, virtual/augmented reality applications~\cite{sigcomm22-cyclops}), 
etc. In general,
transmitter localization is a key technology for location-based services, and an 
improvement in transmitter localization will be very
beneficial to a variety of applications.
In particular, 
in shared spectrum systems~\cite{infocom18-spectrum}, there is a need to guard the shared spectrum
against unauthorized usage which entails detecting and localizing unauthorized transmitters that may
use and/or jam the spectrum illegally.
Classical techniques for transmitter localization involve triangulation~\cite{nsdi13-arraytrack} or fingerprinting techniques~\cite{infocom00-radar} (see~\cite{localization-survey} for a survey).

Advances in quantum technologies have led to the creation of efficient quantum sensors for
radio-frequency (RF) signal detection that are much more sensitive than the classical 
antenna-based RF sensors and are expected to cover the entire 
RF spectrum~\cite{PhysRevApplied.rydberg}. 
E.g., in~\cite{PRL20-qsn},
researchers use some distributed entangled RF-photonic quantum
sensors to estimate the amplitude and phase of a radio
signal, and the estimation variance beats the standard quantum
limit by over 3 dB. 
Thus, QSNs may have a great potential in accurate localization of wireless 
transmitters, which is a problem of great significance in many applications.

\para{Quantum Sensor Model}. 
Impact on a quantum sensor due to a physical phenomenon is typically modeled by an appropriate unitary operator that results in a quantum phase change~\cite{RevModPhys.quantumsensing}. Below, we model the change in quantum phase
of a sensor's state due to an RF received signal. 
Since the RF received signal (and thus the change in quantum phase) depends on the sensor's distance 
from the transmitter, we can use the phase change that occurs 
during the sensing period to localize a RF transmitter.

\softpara{Sensor's Hamiltonian.}
A quantum sensor's Hamiltonian $\hat{H}(t)$ is a sum of two\footnote{The third component 
of control Hamiltonian is chosen to tune the sensor in a controlled way~\cite{RevModPhys.quantumsensing}; we assume $\hat{H}_{control}=0$ in our analysis~\cite{egerstrom}.}
components~\cite{RevModPhys.quantumsensing}:
$$\hat{H}(t) = \hat{H}_0 + \hat{H}_V(t)$$
where $\hat{H}_0$ is the internal Hamiltonian of the system and
$\hat{H}_V(t)$ is the change in the Hamiltonian due to an external signal $V(t)$.
The internal Hamiltonian $\hat{H}_0$ remains fixed and is equal to $E_0 \ket{0}\bra{0} + E_1\ket{1}\bra{1}$, where $E_0$ and $E_1$ are energies corresponding to the 
$\ket{0}$ and $\ket{1}$ states respectively. 
The signal Hamiltonian $\hat{H}_V(t)$ is given by:\footnote{Here, we ignore the 
transverse component of $\hat{H}_V(t)$~\cite{egerstrom}, since, in most sensor applications, the
energy difference $\Delta E = E_1 - E_0$ is much higher than the energy 
changes introduced by the signal $V(t)$~\cite{RevModPhys.quantumsensing}.}
$$\hat{H}_{V}(t) =  -\frac{1}{2}\gamma V_{\parallel}(t)\hat{\sigma}_z$$
where $\sigma_z$ is the Pauli-Z matrix, $V_{\parallel}(t)$ is parallel component of the signal $V(t)$, and $\gamma$ is the coupling of the qubit to the parallel component.
In essence, the above induces a change in the spin in the $z$ axis direction resulting
in a  qubit \emph{phase shift}. Above, 
${V_{\parallel}}(t)$ at the sensor is given by:
$$V_{\parallel}(t) = E \sin(2\pi f  t + \theta)$$
where $E$ is the signal's (electric field) maximum amplitude, $f$ is the signal 
frequency, and $\theta$ is the signal's phase.

\softpara{Evolution Unitary Operator.} Assume at time $t=0$, the quantum state is $\ket{\phi_0}$. Then at time $t=t'$, the state $\ket{\phi_{t'}}$ is,
$$ \ket{\psi_{t'}} = \hat{U}(0, t') \ket{\psi_0} $$
where the time evolution unitary operator $\hat{U}(0, t')$ due to the signal is given by:
\begin{eqnarray*}
\hat{U}(0, t') &=& e^{\frac{i}{\hbar} \int_{0}^{t'} \hat{H}_V(t) dt}  \\
&=& e^{\frac{i}{\hbar} \int_{0}^{t'} (-\frac{1}{2}\gamma V_{\parallel}(t)\hat{\sigma}_z) dt} 
\end{eqnarray*}
where $\hbar=6.626\times 10^{-34} J\cdot s$ is the plank constant.
The unit of coupling $\gamma$ is $J/(V\cdot m^{-1})$, and the unit of $V_{\parallel}(t)$ is $V\cdot m^{-1}$.

\softpara{Phase Shift over a Sensing Time Window.}
Let us represent $\hat{U}(0, t')$ as~\cite{nature21_phase,Zhang_2021}
\begin{equation}
    \hat{U}(0, t') = e^{-\frac{i}{2} \phi  \hat{\sigma}_z} \label{eqn:unitary}
\end{equation}
where the phase shift $\phi  =  \int_{0}^{t'} \frac{\gamma}{\hbar} V_{\parallel}(t) dt$, 
accumulated during the sensing time $[0, t']$ due to the signal $V(t)$ is estimated as follows.
Note that $V_{\parallel}(t)$ is a sinusoidal function---and hence, the phase shift in
one full cycle ($t' =\sfrac{1}{f}$) is zero. To address this, we invert the qubit whenever the sinusoidal function turns from positive to negative using a $\pi$ pulse~\cite{RevModPhys.quantumsensing}.
Thus, the accumulated phase in one cycle $\phi = \int_{0}^{\sfrac{1}{f}} \frac{\gamma}{\hbar} V_{\parallel}(t) dt = \frac{2}{\pi \hbar} \gamma E \frac{1}{f}$.
Since the sensing time $t'$ is expected to be much larger than $1/f$, the phase
shift accumulated during the sensing time $[0, t']$ can be estimated by:
\begin{equation}
    \phi =  \frac{2}{\pi \hbar} \gamma E t'
    \label{eqn:phase_shift}
\end{equation}
Thus, for a fixed sensing time duration, the phase shift in the sensor's quantum state
accumulated due to the signal is proportional to $E$, the signal's maximum amplitude, which is a function of the distance from the transmitter (see~\S\ref{sec:eval}). 

\softpara{Impact on Multiple Quantum Sensors.}
Consider a set of $m$ quantum sensors distributed over an area, with a global $m$-qubit quantum state of $\ket{\psi_0}$. 
Consider a transmitter at a certain specific location in the area. 
Let $\hat{U}_i$ be
the impact on the $i^{th}$ sensor due to the transmitter over a sensing time window. 
Then, the overall impact on the {\em global} quantum state is represented by a \emph{tensor product}  of $m$ individual unitary operators, i.e., $\bigotimes_{i=1}^{m} \hat{U}_i$, and
the evolved global state state is $\bigotimes_{i=1}^{m} \hat{U}_i\ket{\psi_0}$. 

\para{Problem Definition.}
Consider a network of quantum sensors distributed in a geographic area 
and a potential transmitter/intruder in the area.
Let the initial state of the system of quantum sensors network 
be $\ket{\psi_{0}}$. 
As described above, due to the transmission from the intruder, the quantum state evolves to $\ket{\psi_{t'}} = \hat{U} \ket{\psi_{0}}$ over a period of time $t'$.
The transmitter localization problem is to determine the location of the transmitter based on
the evolved quantum state $\ket{\psi_{t'}}$.

\para{Related Work.}
Radio transmitter localization using a set of sensors/receivers has been
widely studied~\cite{localization-guide, localization-survey, time-19}. 
Localization methods can be roughly classified into two types: geometry-based and fingerprinting-based.
The geometry-based method includes multilateration (by measuring time-of-flight between the transmitter and multiple sensors) or triangulation (by measuring angle-of-arrival (AoA) of the transmitter at multiple sensors)~\cite{nsdi13-arraytrack}.
The fingerprinting-based method~\cite{infocom00-radar} has a training stage that records
the signal fingerprint for certain locations; Localization is then achieved by
matching the real-time signal to the recorded fingerprints. 
Here, a fingerprint for a transmitter location may be a vector of received signal strengths (RSS)~\cite{localization-guide} at the sensors.  
Localization of simultaneously-active multiple transmitters is more challenging,
and has been addressed in recent works~\cite{wowmom, ipsn20-mtl, pmc22-deepmtlpro}.

Recently, there have been some works that have used quantum technology to investigate
intruder/transmitter localization related problems. E.g.,~\cite{lcn22-qloc} develops a scheme
to improve the {\em size} of the fingerprints used in the above-described 
fingerprinting approach, by encoding classical sensor data into qubits 
through quantum amplitude encoding.
In addition,~\cite{PR22-quantum_positioning} derives analytical equations to compute AoA of an
incoming RF signal using four entangled distributed quantum sensors, without any evaluations.~\cite{qsn-detection} proposes a quantum sensor network 
using Mach-Zehnder interferometers to detect (not localize) 
intruders for surveillance purposes.
Finally,~\cite{PhysRevA.quantum_sensors, qsn-acm-23} investigate the optimization of initial state in discrete-outcome quantum sensor networks and show that an entangled initial state yields
higher measurement accuracy in some applications.

\softpara{Parameter Estimation using Quantum Sensors.}
Prior works on parameter estimation using quantum sensors
include: estimation of single~\cite{Giovannetti_2011} 
or multiple independent parameters~\cite{Proctor_2018}, estimation of a single linear function over parameters~\cite{Altenburg_2019}, and estimation of multiple linear functions~\cite{Rubio_2020}. Our transmitter localization problem can be looked upon
as a novel single parameter (TX location) estimation problem based on sensor measurements that are functions (based on signal propagation model and distance) of the parameter being estimated.

\section{\bf Quantum State Discrimination  Based Algorithms}
\label{sec:povm}

\para{Quantum State Discrimination (QSD).} 
Given a quantum state $\ket{\phi}$ that is known to be equal to one of the 
states (known as \emph{target states}) in the set 
$\{\ket{\phi_1}, \ket{\phi_2}, \ldots, \ket{\phi_n} \}$,
the quantum state discrimination (QSD) problem 
is to determine which state $\ket{\phi}$ really is.
In general, each target state $\ket{\phi_i}$ may be associated with
a prior probability $q_i$; in this paper, we assume uniform
prior.
The QSD problem is typically solved using a series of measurements or 
a single measurement---as defined below.
It is known that if the target states $\{\ket{\phi_i}\}$  are not mutually orthogonal, 
then there is no quantum measurement capable of perfectly (without error)
distinguishing the states.
Thus, a QSD solution may give an erroneous answer---i.e., guess the
state to be in $\ket{\phi_i}$ when the state is really in $\ket{\phi_j}$ for 
some $i \neq j$. Thus, a QSD solution is associated with an overall {\em probability
of error} (PoE), and the optimization goal of the QSD problem is to determine the
measurement (or a sequence of measurements) that minimizes the PoE. 
We note that in our developed schemes, we don't actually solve the QSD 
problems that
arise due to the impracticality of implementing the general POVMs, as discussed
later; instead,
we just use the standard POVM known as pretty good measurement (PGM).

\softpara{General Measurements.} 
A general measurement~\cite{qcqi-book} is defined by matrices $M_1, M_2, \ldots, M_n$ such that
$\sum_i M_i^{\dagger}M_i = I$ where $M_i^{\dagger}$ is the conjugate transpose of
$M_i$. If this general measurement is carried out on a pure state,
we see the outcome ``$i$'' with 
probability $p(i) = \bra{\phi}M_i^{\dagger}M_i\ket{\phi}$. Thus, if we associate
the outcome ``$i$'' with the given state $\ket{\phi}$ being in the target state 
$\ket{\phi_i}$, the probability of error (PoE) for the given measurement $\{M_i\}$
is given by $\sum_i \sum_{j \neq i} \bra{\phi_i}M_j^{\dagger}M_j\ket{\phi_i}$.

If we are only interested in the probability of outcomes (as in our context), the above
general measurement can also be represented by the set of positive semi-definite 
matrices (PSD) $\{E_i = M_i^{\dagger}M_i\}$ where $\sum_{i} E_i = I$. This representation 
is called positive-operator valued measure (POVM); in this paper, we use this 
representation of measurement for simplicity. 


\para{Core Idea: TX Localization as QSD.}
Consider a geographic area where a transmitter can be at a set of potential locations $\{l_1, l_2, \ldots, l_n\}$. 
For simplicity, let us assume that the transmission power is constant.
Let the initial state of the quantum system, composed of say $m$ distributed quantum sensors,
be $\ket{\psi_0}$. 
When the transmitter $T$ is at a location $l_i$, let the impact of the $T$'s transmission from location $l_i$ evolve the overall state of the quantum system to $\ket{\psi_i}$ 
based on the model described in the previous section.
Now, consider the set of target states $\{\ket{\psi_1}, \ket{\psi_2}, \ldots, \ket{\psi_n}\}$ corresponding to the set of potential locations of the transmitter. Then, localizing the transmitters, i.e., determining the location $l_i$ from where the transmission occurred, is
tantamount to solving the QSD problem with the target states $\{\ket{\psi_i}\}$.
Thus, determining the state of the quantum system yields
the transmitter location.

\softpara{Selection of Initial State and Measurement.} 
In the above context, our goal is to select an initial state $\ket{\psi_0}$ and the 
POVM measurement (i.e., PSD matrices $\{E_1, \ldots, E_n\}$, one for each potential outcome/location) such that the overall \eat{probability of error} PoE is minimized --- for a given setting of transmitter location, quantum sensors, and signal propagation model.
The optimization problem of selecting an optimal combination of initial state and POVM
in our context is beyond the scope of this work. Here, we use a non-entangled uniform superposition
pure initial state $\ket{\psi_0} = \sum_{i=0}^{2^m-1} \frac{1}{\sqrt{2^m}} \ket{i} $. 
For a given initial state and target states, determining an optimal POVM can be shown to
be a convex optimization problem and can be solved using an appropriate semi-definite program (SDP)~\cite{semidefinite}. However, due to scalability challenges in solving the
SDP, whose size is exponential in the number of quantum sensors involved, 
in this paper, we use
a well-known measurement known as {\em pretty-good-measurement} (PGM) which is known to
perform well in general settings~\cite{prettygood}. The PGM POVM is given by: 
\begin{equation}
    E_i = q_i {\rho}^{-1/2} \rho_i {\rho}^{-1/2}
    \label{eqn:pgm}
\end{equation}
where $q_i$ is the prior probability and $\rho_i = \ket{\psi_i}\bra{\psi_i}$ is the {\em density matrix} of the $i^{th}$ target state $\psi_i$, and  $\rho = \sum_{i} q_i \rho_i$. 

\begin{figure}
    \centering
    \includegraphics[width=0.44\textwidth]{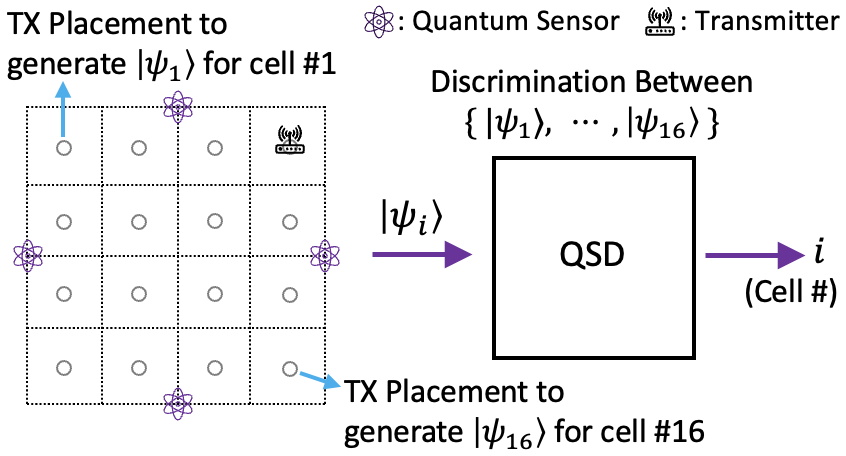}
    \caption{\povmone Scheme.}
    \label{fig:onelevel}
    \vspace{-0.2in}
\end{figure}

\begin{figure*}[t]
    \centering
    \includegraphics[width=0.95\textwidth]{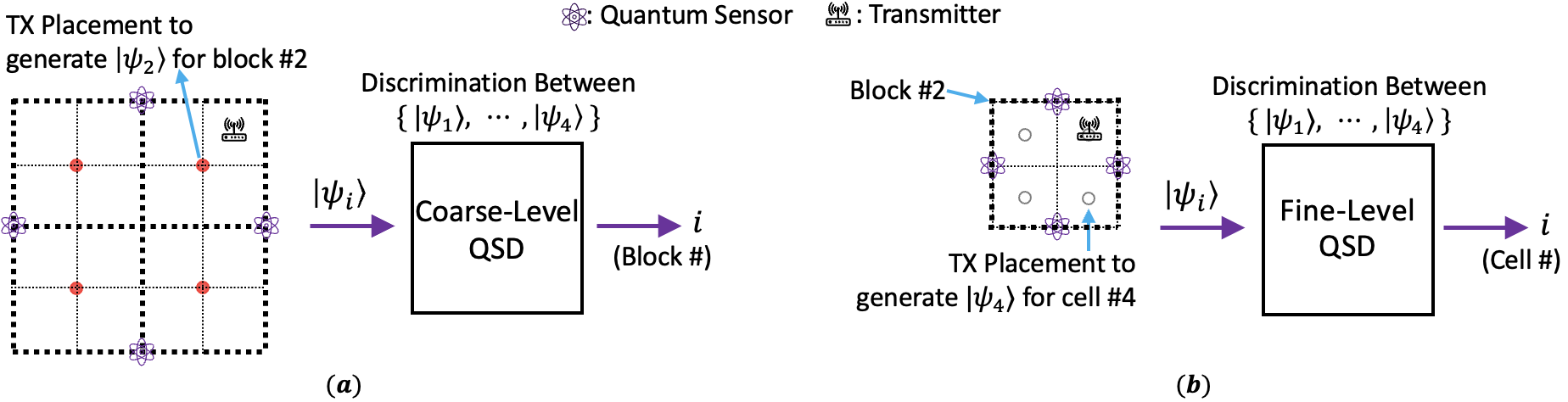}
    \caption{\povm Scheme. (a) Coarse-level localization phase, and (b) Fine-level localization phase.}
    \label{fig:twolevel}
    \vspace{-0.2in}
\end{figure*}

\para{Basic \povmone Scheme; Key Challenges.}
The above-described methodology is essentially our basic \povmone localization scheme, see Fig.~\ref{fig:onelevel}. 
That is, the \povmone scheme localizes the transmitter by first determining the set 
of target states $\{\ket{\psi_1}, \ket{\psi_2}, \ldots, \ket{\psi_n}\}$ corresponding 
to the centers of the cells (a set of transmitter locations), and then, localizes the 
transmitter
in real-time by performing QSD over the evolved quantum state using PGM measurement.
Note that we use only the cells' centers to generate the target states, and also that
the predicted location
of the transmitter is always a cell's center in the QSD-based schemes, since the QSD-based
schemes are fundamentally classification of the transmitter location into cells.
{\em However, during evaluation, the actual
location of the transmitter can be anywhere in the area}---presumably, non-center locations 
of the transmitter may incur higher localization errors.

The key challenges in the \povmone scheme are twofold: (i) It is likely to incur a high probability
of error due to a large number of target states (equal to the number of potential 
transmitter locations). (ii) A global POVM measurement over a large number of sensors can be
difficult to implement in practice~\cite{pra19-povm}; even ignoring the communication cost of teleporting the 
qubits to a central location, the main challenge arises due to the complexity of the circuit or
hardware required  to implement a POVM over a large number of qubit states.
We address these challenges by designing a two-level localization scheme as described below; 
in the following section, we further address the above challenges by designing non-QSD based
schemes.


\para{\povm Scheme}. 
\povm solves the above-mentioned challenges by localizing the transmitter by using two levels
of POVMs, with each POVM requiring a measurement over a much fewer number of sensors and with a much fewer number of 
possible target states. 
We discretize the given area into a grid; each unit of the grid is called a \emph{cell}. 
A \emph{block} is a group of neighboring cells that form a rectangle. 
In Fig.~\ref{fig:twolevel} (a), a grid has $4\times 4$ cells and $2\times 2$ blocks. 
The thick dotted lines depict the blocks while the non-thick dotted lines depict the cells.
In general, for a $N \times N$ grid with $N^2$ cells, we construct blocks by dividing the entire grid
into $\sqrt{N} \times \sqrt{N}$ blocks --- yielding $N$ blocks in the whole area, with each block comprised of $ \sqrt{N} \times \sqrt{N} = N$ cells. Without loss of generality, we assume $\sqrt{N}$ to be an integer in our discussion.
The basic idea of the \povm scheme
is to localize the transmitter in two stages: first, localize the transmitter at a block level (Fig.~\ref{fig:twolevel} (a)); and then, within that block, localize the transmitter at the cell level (Fig.~\ref{fig:twolevel} (b)). The sensors, target states, and POVMs used for localization at these two stages are different. Such a two-stage localization scheme naturally addresses the above-mentioned challenges by reducing both the number of sensors as well as target states required at 
each stage. We describe the scheme in more detail below.

\softpara{Coarse-Level Localization.} 
The coarse level concerns localizing the transmitter at the block level, and is done based on {\em coarse-level sensors} deployed over the entire given area. The target states for the coarse-level QSD/localization are the states corresponding to the location at the center of each block in the given area. As mentioned above, since the number of blocks is $N$, the number of target states for the Coarse-Level localization is $N$. The POVM measurement for the coarse-level localization is constructed using 
Eqn.~\ref{eqn:pgm} for the PGM measurement over the target states derived from the impact of 
the transmitter at coarse-level discrete locations (i.e., the center of the blocks) on the coarse-level sensors.
Note that in reality, the transmitter is likely not at the center of the blocks---but, we stipulate that a block's center is a reasonable representative of the actual locations of the transmitter in that block.
More formally, let $\{L_1, L_2, \ldots, L_N\}$ denote the centers of the blocks in the
area, and $S$ be the coarse-level sensors. Let $\hat{U}_i$ denote the impact on $S$ when the transmitter is at location $L_i$. Then, the target states for the coarse-level localization 
are $\{\hat{U}_i \ket{\psi_0}\}$ where $\ket{\psi_0}$ is the initial state of $S$. 
These target states are used to determine the POVM measurements as per Eqn.~\ref{eqn:pgm}, and thus, determine the block.

\softpara{Fine-Level Localization.}
Once the transmitter has been localized within a block $B$ via coarse-level localization, the 
transmitter is then localized at a cell level within $B$. For fine-level localization, each block $B$ has a set of fine-level sensors $S(B)$ deployed within $B$ (which need not be 
disjoint from the coarse-level sensors). 
The target states for fine-level localization within $B$ correspond 
to the potential locations of the transmitter within $B$ which are the centers
of the cells within $B$, see Fig.~\ref{fig:twolevel} (b), and is derived from the impact of the transmitter's signal at
the fine-level sensors $S(B)$. 
Note that at the fine-level localization phase, only the sensors $S(B)$ where $B$ is the block selected in the previous coarse-level localization are involved. Note that $S(B_1)$ and $S(B_2)$ from two different blocks need not be disjoint. 
\eat{since sensors only their common border could be involved in fine-level localization within both blocks.}
More formally, let $\{l_1, l_2, \ldots, l_N\}$ denote the centers of the cells in the
block $B$ selected by the coarse-level localization phase, and $S(B)$ be the fine-level sensors. Let $\hat{U}_i$ denote the impact on $S(B)$ when the transmitter is at location $l_i$. Then, the target states for the fine level localization 
are $\{\hat{U}_i \ket{\psi_0}\}$ where $\ket{\psi_0}$ is the initial state of $S(B)$. These target states
are used to determine the POVM measurement as per Eqn.~\ref{eqn:pgm}, and thus, determine
the cell within the block $B$, which is the TX location. 
As mentioned before in the one-level scheme, we note that, during evaluation, the location of the transmitter can be anywhere in 
the area, even though we have only use the cells' centers to generate the target states.

\softpara{Multi-shot Discrimination.}
The quantum measurement is intrinsically probabilistic and the single-shot discrimination can incur a high probability of error. One way to reduce this probability of error is to repeat the 
discrimination process many times and pick the most frequent measurement outcome. 
Such repeated measurements are commonly done in quantum sensing~\cite{RevModPhys.quantumsensing} and computing~\cite{Shor_1997}.
In our context, the repetitions are done while the transmitter remains fixed.



\eat{This repetition is applied to the \povmone, the coarse/fine level of \povm, and also the \povmpro in the following paragraphs.}

\section{\bf Parameterized Quantum Circuit Based Localization}
\label{sec:pqc-loc}

\begin{figure*}
    \centering
    \includegraphics[width=0.98\textwidth]{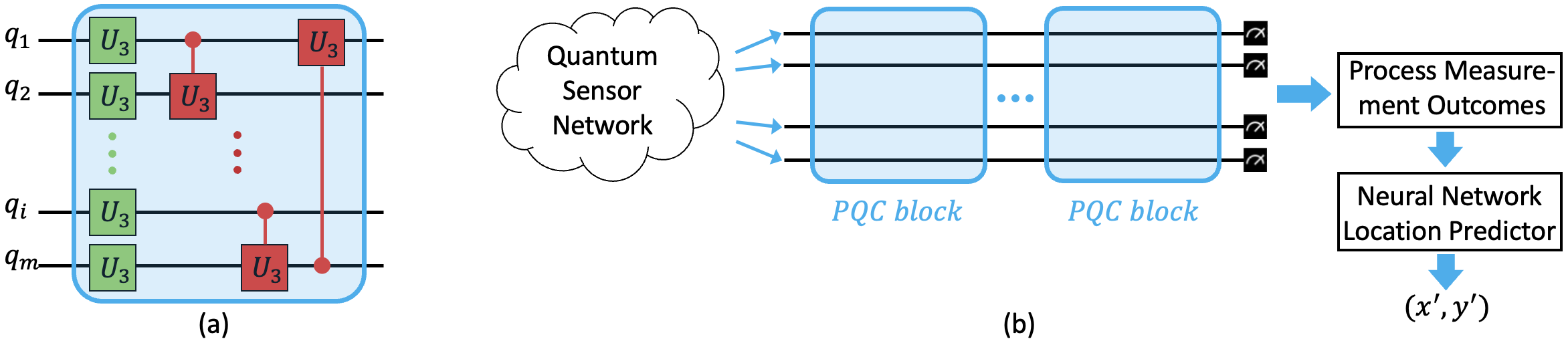}
    \caption{(a) Our parameterized quantum circuit (PQC) block, for the general case of $m$ qubits. It contains $m$ number of $U_3$ gates and $m$ number of $CU_3$ gates.
    (b) The hybrid quantum-classical circuit to localize a transmitter. It consists of multiple PQC blocks, followed by classical processing of measurements, and finally, a neural network-based location predictor. We use only four blocks of PQCs in our hybrid circuit.
    }
    \label{fig:pqc}
\end{figure*}

\para{Motivation.} The QSD-based method discussed in the previous section has a solid mathematical foundation, but its practical implementation is non-trivial and even infeasible for a large sensor network and/or a large number of potential transmitter locations. 
In particular, the POVM measurement operator (derived from the QSD problem or corresponding to the pretty-good measurement)
can be infeasible to implement for a 
large number of outcomes/locations. 
The issue is somewhat mitigated by using a two-level approach as described above, but the PoE (probability of error) 
in the first (coarser) level can still be high due to imperfect training (as we assign a single outcome to a {\em set} of
target locations). 
A potential approach to address the above challenge is to ``translate'' a given POVM into an appropriate quantum circuit comprised of quantum gates and simple
measurements (e.g., projective and/or computational basis)~\cite{pra19-povm}. E.g.,~\cite{qcompiler} presents a technique to convert POVM operators to such quantum circuits. 
However, the computational time incurred in translating a POVM operator into a quantum circuit 
is exponential to the number of qubits and is thus infeasible. In addition, the translated quantum circuits 
are also sub-optimal in terms of the number of CNOT gates used~\cite{qcompiler}.
Also, note that the POVM computed in our QSD-based method is sub-optimal to begin with.

In this section, we develop a machine learning (ML) technique to actually learn a quantum circuit that represents the processing and measurement protocol needed to localize the transmitter from the evolved quantum state.
The learned quantum circuit model maps the global evolved quantum state to the transmitter location.
In essence, we avoid computing the POVM (from QSD, or using the pretty-good measurement) altogether (and thus, also avoid the challenge of translating it to a quantum circuit), and instead learn the required quantum circuit representing the measurement protocol.
To facilitate learning the quantum circuit, we use an appropriate parameterized quantum circuit (PQC) and learn 
its parameters---as in~\cite{PhysRevResearch.qsd-qnn} 
wherein a POVM is trained using parameterized quantum circuits.
Our PQC-based localization method based on the above insights is described
below. We start with a brief introduction to PQCs.

\para{Parameterized Quantum Circuits (PQC)}.
Parameterized quantum circuits have emerged as a powerful tool in quantum computing~\cite{Benedetti_2019}, providing an adaptable framework for tackling diverse computational tasks. 
Parameterized quantum circuits (PQCs) can be regarded as machine learning models with remarkable expressive power; just like classical ML models, PQC circuits/models are trained to perform data-driven tasks. 
PQCs offer several advantages over fixed quantum circuits~\cite{Benedetti_2019, quantumnas2022,quantumnat2022}, including:
\begin{enumerate}
    \item Adaptability. The parameters in PQCs can be adjusted to tailor the circuit for a specific problem, allowing a single circuit structure to be repurposed for various tasks.
    \item Trainability. PQCs can be trained using classical optimization algorithms to solve optimization problems and machine learning tasks, making them a vital component of hybrid quantum-classical algorithms. 
    \item Noise Resilience. PQCs can be more robust against noise and errors in near-term quantum devices, as they allow shorter-depth circuits that reduce the impact of errors. 
\end{enumerate}

In essence, PQCs are quantum circuits comprised of parameterized gates and measurements.
Commonly used parameterized gates in PQCs include rotation gates $R_x(\theta),R_y(\theta),R_z(\theta)$ which represents rotating about the $X,Y,Z$ axis respectively with angle $\theta$.
A more versatile gate is the $U_3(\theta, \phi, \lambda)$, which can be used to generate any single-qubit operation by setting the appropriate values for the parameters;
$U_3$ gate can be decomposed into simpler $R_x, R_y, R_z$ gates.
The parameterized $CU_3(\theta, \phi, \lambda)$ gate is the controlled 
version of the $U_3$ gate; it applies the $U_3$ gate only when the control qubit 
is in the $\ket{1}$ state.
We use a combination of $U_3$ and $CU_3$ gates in our parameterized quantum circuits. 

\para{PQC-based Localization Method}.
At a high level, in our PQC-based localization scheme, the QSN data is fed into a trained hybrid quantum-classical model, which represents the overall measurement strategy and thus outputs the transmitter location. 
Our hybrid quantum-classical model (see Fig.~\ref{fig:pqc}(b)) consists
of the following three components.
(i) Parameterized Quantum Circuit (PQC), (ii) Processing the measurement outcomes,
(iii) Neural network location predictor, to convert the processed measurement outcomes to the transmitter location.
We describe each of the above components below.

\softpara{1. Parameterized Quantum Circuit (PQC) Design.}
The parameterized quantum circuit can be designed in many ways. 
We design our PQC component based on some common PQC-design patterns~\cite{liang2023unleashing, ICCAD21-wang} used in prior works.
For example, in~\cite{lloyd2020quantum}, a block of 
PQC contains one layer of $ZZ$ gates 
and one layer of $R_y$ gates.
In~\cite{McClean_2018}, a block of PQC contains one layer each
of $R_x, R_y, R_z, CZ$ gates.
In our scheme, the quantum circuit is composed of blocks, and 
each block is a combination of $U_3, CU_3$ gates;
we use these two gates in our design as they form a universal
gate set and are widely used in PQC circuits.
Circuits consisting $U_3$ and $CU_3$ gates have a high expressive power as each gate has three trainable parameters.
In particular, given $N$ number of input qubits, a block consists of 
$N$ number of $U_3$ and $N$ number of $CU_3$ gates. 
See Fig.~\ref{fig:pqc}.
In a block, each input qubit is first  
operated on by the unary $U_3$ gate in parallel,
forming a layer of $U_3$ gates. 
Then, there is a series of $CU_3$ gates following a ring connection pattern, 
i.e., each $CU_3$ is executed over two ``consecutive'' qubits (with the first being
the control qubit) except for the
last $CU_3$ gate which is over the last and the first qubit. 
Thus, a single block has a circuit depth of $N+1$.
The overall PQC may have a series of above blocks---the expressive power of the model increases monotonically with the increase in the number of blocks.
In our evaluations (\S\ref{sec:eval}), we used four blocks as we observe that four blocks provide good performance while having a modest circuit depth.
After the blocks, the PQC ends with the measurement on the standard computational basis, 
i.e., the Pauli Z basis.

\softpara{2. Process Measurement Outcomes.} 
As in the QSD-based schemes, we will use the PQC to make repeated measurements.
To use the repeated measurements effectively for location prediction, we characterize
the set of repeated measurement results by expectation values, one for each qubit.
In particular, we compute the expectation value 
$\langle Z \rangle$ of the Pauli Z operator 
(which represents
the measurement in the computational basis), and feed as input to a neural network for 
final location prediction as described below. 
We note that, for a quantum state $\ket{\psi} = \alpha \ket{0} + \beta \ket{1}$, the expectation value 
$\langle Z \rangle$ of the Pauli Z operator is given by $\bra{\psi} Z \ket{\psi} = |\alpha|^{2} - |\beta|^{2}$.


\softpara{3. Neural-Network to Predict Location.}
We consider two variants of our neural network predictor: (i) Classifier variant. which outputs a class/label corresponding to the cell where the TX is located, and thus, predicts the location to be the cell's center. (ii) 
Regression variant, that outputs locations as the $x$ and $y$ coordinates.

\begin{figure}
    \centering
    \includegraphics[width=0.35\textwidth]{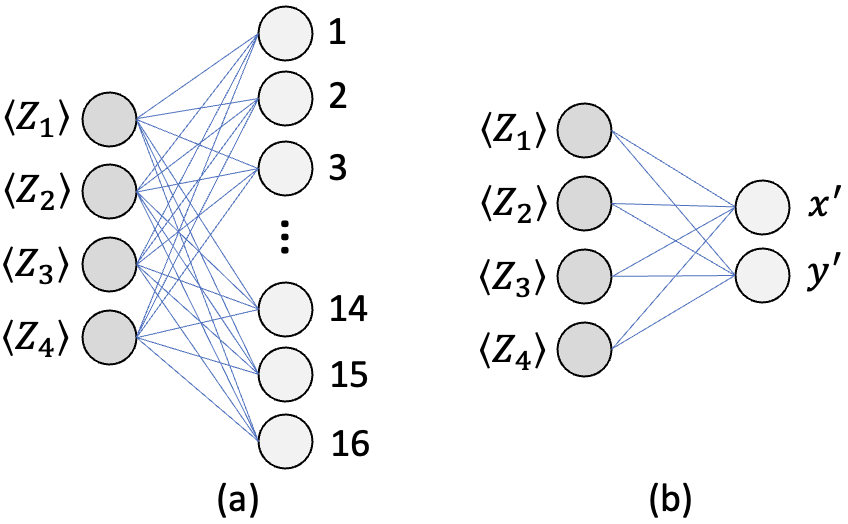}
    \caption{Neural network (a fully connected layer) for 4 quantum sensors to predict the location from processed measurements. (a) Classifier Variant, (b) Regression Variant.}
    \label{fig:fclayer}
\end{figure}

{\em Classifier Variant.}
Our overall circuit with the Classifier component for the location prediction essentially equates to a circuit for quantum state discrimination (QSD), as the 
QSD problem also outputs a finite number of discrete outcomes.
For the Classifier Variant, we use a simple neural network with only an input layer and an output later, having no hidden layers---i.e., a single 
fully connected layer.
The input neurons are the expectation values of the Pauli Z operator from the measurements as described above, and the output neurons represent the cell labels. See Fig.\ref{fig:fclayer}(a), which shows the fully connected layer for a network of 4 quantum sensors deployed in a $4\times 4$ grid with 16 cells.

{\em Regression Variant.}
The Classifier Variant outputs locations in a discrete space---which is fundamentally sub-optimal if the transmitter can be anywhere in the 2D space. 
To output the predicted location in the continuous 2D space, we use a Regression Variant that outputs the location as an $(x,y)$ point.
For the setting wherein the transmitter may be located anywhere in the 
2D space, the Regression Variant should have a smaller localization error.
Fig.\ref{fig:fclayer}(b) shows the fully connected layer for the Regression Variant; the number of output neurons is always two, i.e., a $X$ coordinate and a $Y$ coordinate.

\softpara{4. Loss Function.}
During training, the gradient of the loss function is back-propagated through the neural network and the quantum circuit parts, so that the parameters within these parts can be appropriately updated. 
The loss functions used for the Classifier Variant and the Regression Variant are different; for the Classifier variant, we use the cross-entropy loss function while for the Regression variant, we use the mean square error loss function.

\para{\pqcone and \pqctwo Schemes.} 
The above-described hybrid quantum-classical model is essentially our \pqcone localization scheme.
By using \pqcone as a building block and using the same two-level (coarse, fine) idea described in~\S\ref{sec:povm}, we design the \pqctwo, corresponding to the two-level QSD-based schemes described in the previous section.
At the first coarse level, the output of the ``coarse-level \pqcone'' will determine the block the transmitter is in.
Then at the second fine level, the output of the ``fine-level \pqcone'' tied to the block determined by the coarse level is the final location output.
The PQC-based schemes essentially use the trained circuit in lieu of the POVM used in the QSD-based schemes. 
The PQC-based schemes can be used with either the Classifier or the Regression variant
for the last predictor component.

\section{\bf Evaluation}
\label{sec:eval}

In this section, we evaluate our developed schemes.
We make two observations, which are as expected:  (1) Performance of two-level methods is better than one-level methods in general, and (2) Performance of the PQC-based methods is superior to the QSD-based methods. In summary, our schemes are able to achieve meter-level (1-5m) localization accuracy, and near-perfect (99-100\%) classification accuracy in the case of discrete locations.


\subsection{Evaluation Settings}

\para{Algorithms Evaluated.}
We evaluate four algorithms: \povmone, \povm, \pqcone, \pqctwo. As the name implies, \povmone and \pqcone are one-level methods, while \povm and \pqctwo are two-level methods.
Similarly, \povmone and \povm are QSD-based methods, while \pqcone and \pqctwo are PQC-based methods. We use the Regression Variant in our PQC-based methods by default, since in our default setting the transmitter can be anywhere in the 2D space.
Our code\footnote{\url{https://github.com/caitaozhan/QuantumLocalization}} is written in Python and uses Numpy and Scipy libraries to perform matrix-related operations.


\para{QSD-based Method Implementation.}
To implement the QSD-based methods, we first determine the target states which are then used to construct the pretty-good-measurement POVM via Eqn.~\ref{eqn:pgm}.
To localize a transmitter, we first compute the evolved state and then, use the POVM to determine the target state or the TX location. 
This is done repeatedly as described in Section~\ref{sec:povm}, and in two levels (coarse, fine) depending on the localization scheme. 
The target states and evolved states are both generated using the sensor model described
in Section~\ref{sec:problem}, i.e., using Eqns.~\ref{eqn:unitary}-\ref{eqn:phase_shift}, with the electric field strength ($E$) and phase shift range modeled as below. 

\para{Generating Sensor Readings.}
The crux of determining target states, computing the evolved states, and simulating the training datasets is to compute the phase shift picked up by the quantum sensors due to the signal during the sensing process.
In Eqn.~\ref{eqn:phase_shift}, we modeled the phase shift as a function of the electric field strength and the sensing time.
Thus, we need a model for the electric field strength.
In free space, the electric field strength produced by a transmitter with an isotropic radiator can be approximated as~\cite{e-field-wiki}
$$E = \frac{\sqrt{30 \cdot P}}{d} \times (1 + noise) $$
where $E$ is the electric field strength in $V\cdot m^{-1}$, $P$ is the transmitter power output in $W$ (watt), and $d$ is the distance from the radiator in $m$.
Since in most quantum sensing applications, the signal to be sensed are weak signals, here we assume the power of the transmitter $P=0.1 \mu W$.
Ideally, the strength of the electric field is inverse to the distance between the transmitter and the sensor.
But in reality, the relationship is more complicated.
So, we add a random uniform variable $ noise\in[-0.05, 0.05]$ to incorporate reality in a simple way.
The target states and thus the POVMs are computed assuming zero noise during training, while during localization, the signal received is assumed to contain noise. 
The simulated datasets for PQC-based methods are assumed to contain noise too.

\softpara{Range of Phase Shift $\phi$.}
We set the sensing time $t'$ to 1 millisecond\footnote{In principle, the sensing time period must be smaller than the decoherence time, which varies across different quantum technologies.}. 
As mentioned later, our grid cells are $10m \times 10m$, and we assume 5 meters to be the minimum
distance allowed between a transmitter and a quantum sensor.
Thus, 
we choose 
the coupling constant $\gamma$ to be such that a quantum sensor at 5 meters away 
from the transmitter accumulates a phase shift of $2\pi$ during the sensing time $t'$; this entails
that the maximum phase shift is $2\pi$ and the minimum phase shift is as low as $0$ 
(when the sensor is very far away from the transmitter).

\para{PQC-Based Methods Implementation and Training.}
Different than the QSD-based methods, the PQC-Based methods involve quantum circuits.
We use the publicly available TorchQuantum~\cite{quantumnas2022} library to implement and train the parameterized hybrid circuits.
TorchQuantum's classes are inherited from a core class of PyTorch~\cite{pytorch}, which is used to implement the neural network predictor. 
Thanks to PyTorch, we are able to train the PQCs fast on a GPU.
We use the Adam optimizer and train for 80 epochs for both \pqcone and \pqctwo methods.

The sensor readings are also used as the sensor data to train the PQC-based hybrid circuit models. 
Essentially, for a fixed initial global state of the sensors (say, $\psi$), each sample consists of the quantum state received from the quantum sensor network (input feature) and the location of the transmitter (ground truth target). 
More formally, each sample is of the kind: 
$(\bigotimes_{i=1}^{m} \hat{U}_i\ket{\psi}, L)$, where $\hat{U}_i$ is the evolution unitary operator for the $i^{th}$ quantum sensor (as per \S\ref{sec:problem} and above paragraphs), $\ket{\psi}$ is the uniform superposition initial state, and $L$ is the location of the transmitter in the field in all scenarios except for one, i.e., $L$ is the block number for samples used to train a ``coarse-level \pqcone'' in the \pqctwo Classifier Variant. 
We use one hundred training examples/samples for each cell, with the transmitter's location randomly scattered over the cell. 
For example, consider a $4\times4$ grid with a block length of 2.
The training dataset for \pqcone will have $16\times100=1600$ samples. 
And \pqctwo will have $16\times100=1600$ samples in the first level to train a ``coarse-level \pqcone'', and 
$4\times400=1600$ samples in the second level to train 4 blocks each requiring
$4\times100=400$ to train a ``fine-level \pqcone''.
Thus, there are a total of 3600 training samples used to train 5 models in a \pqctwo method.

\para{Quantum Sensor Deployment.} 
We deploy sensors uniformly over the area; for the \povm and  \pqctwo schemes, we deploy the fine-level sensors along the block borders so that the sensors can be used by the two neighbor blocks, i.e. fine-level sensors for the blocks are not disjoint.
We use a maximum of 8 quantum sensors for any single QSD instance---since the memory and computing requirements for storing and implementing a POVM become prohibitive beyond that. E.g., a POVM for 256 target-states over 12 sensors requires 69 GB of main memory storage.\footnote{We need $256$ matrices each of size $2^{12} \times 2^{12}$, with each matrix element being a complex number requiring 16 bytes.} 
The PQC-based methods have a bottleneck on the number of sensors due to POVM considerations, but 
we are still limited in practice nevertheless due to training time and GPU memory; thus, we 
use a maximum of 16 sensors in the first level or in any block of the second level. This limits
the training time to at most several hours and GPU memory requirements to 8-16 GB. 
We discuss more details on number of sensors used at various levels and blocks, below.
Finally, each grid cell is of size $10m \times 10m$ in all settings, and the transmitter 
can be anywhere in the given area except that the minimum distance between any sensor
and transmitter is 5m.

\softpara{Two-Level Schemes: Blocks and Sensors Used.}
As described in \S\ref{sec:povm}, for a grid $N \times N$, if $N$ as a perfect square, 
the grid is divided into $\sqrt{N} \times \sqrt{N}$ blocks---with the first-level localizing the transmitter into one of the blocks, and the second-level localizing
the transmitter into a cell within the block.
However, in this section, to get a better insight into the performance trends, in this section, we have also considered  $N$ values that are not perfect squares. 
For such $N$ values, we have determined block sizes as integers close to the $\sqrt{N}$; e.g., for a $12\times12$ grid, we divided the grid into $4 \times 4$ blocks each of $3 \times 3$ cells.
In terms of the number of sensors at each level---we use up to 16 sensors in the first-level of localization, but in the second-level we always use exactly 4 sensors per block irrespective of the block/grid size.

\para{Performance Metrics.} 
We use the {\tt Localization error} ($\lerr$, in meters) as the main metric to evaluate our localization schemes. $\lerr$ is defined as the distance between the actual location of the transmitter and the predicated location. 
In all plots except the CDF plots, average $\lerr$ refers to the
{\em average} localization error over many TX locations; 
in the CDF plots, the distribution is over many TX locations.

\subsection{Evaluation Results}
In our evaluation, we evaluate the performance of our proposed four localization algorithms' performances for varying  grid size and number of quantum sensors.
Note that, for one-level algorithms, the number of sensors is the total number of sensors used, while for the two-level algorithms, the number of sensors parameter is the number of sensors used in the first/coarse level (recall that, in the second level, we use only 4 sensors for each block). 
\begin{figure}[h]
    \centering
    \includegraphics[width=0.38\textwidth]{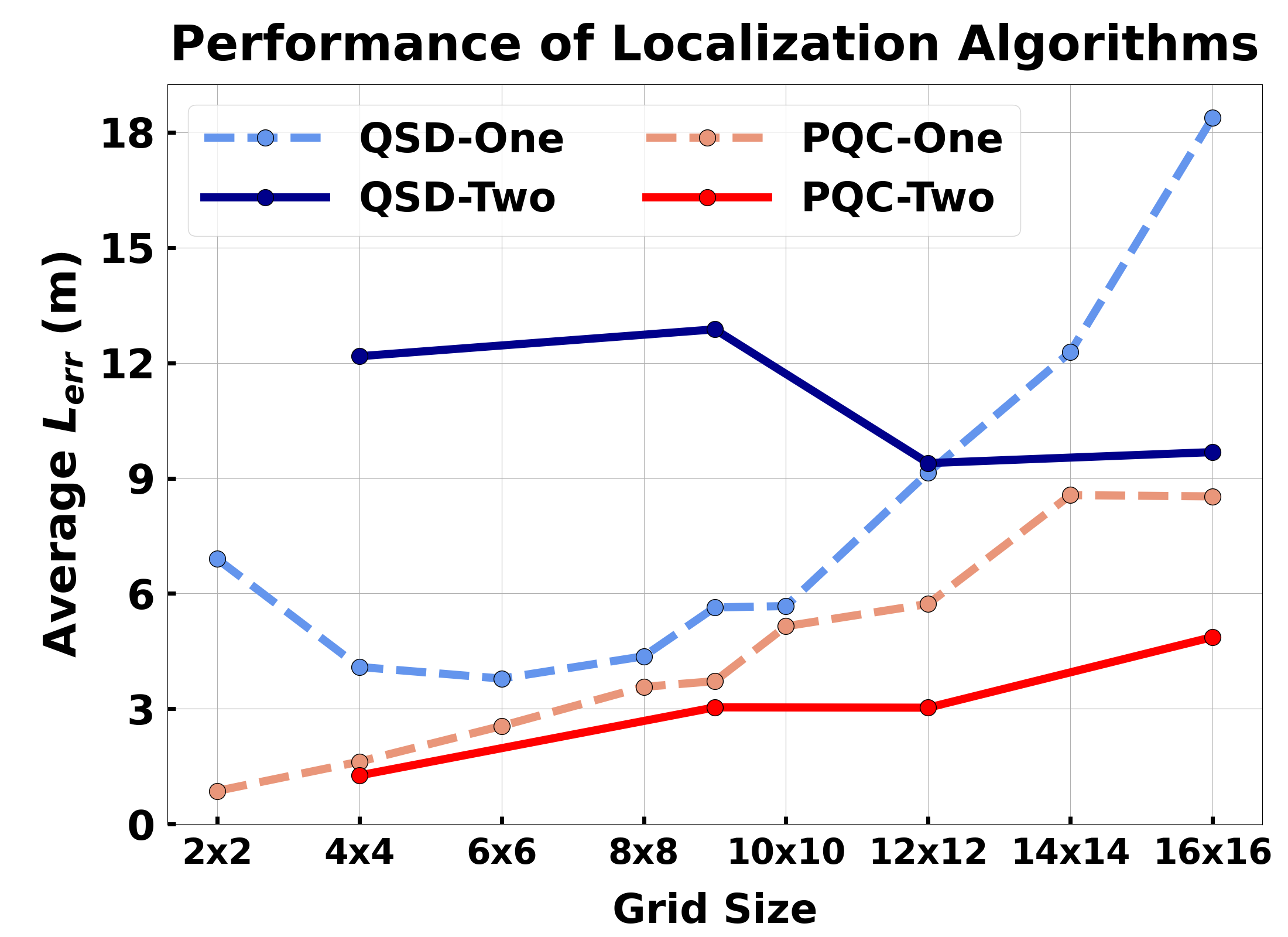}
    \caption{The performance of \povmone, \povm, \pqcone, \pqctwo for varying grid size and 8 quantum sensors.}
    \label{fig:continuous.varygrid}
\end{figure}

\para{Varying Grid Size.} Fig.~\ref{fig:continuous.varygrid} shows the performance of all four algorithms with varying grid sizes when the number of quantum sensors is eight. 
We observe that the PQC-based methods have lower localization error than the QSD-based methods, and
the two-level schemes generally perform better than one-schemes---except that the \povm performs worse than \povmone for smaller grid sizes.\footnote{This is likely because the QSD problem at the first/coarse level has a high error. 
The high error here is due to all cells being at the border edges, making the quantum state discrimination hard.
The neighboring two cells across the border of two blocks are close, thus hard to determine which block the cell is in.}
The results show the power of a well-trained parameterized hybrid circuit and the effectiveness of two-level schemes. 
More specifically, we observe that for a $16\times16$ grid, the average $\lerr$ of \pqctwo is $4.9 m$, which is almost half of the average $\lerr$ of \pqcone at $8.5m$.
Similarly, the $\lerr$ of \povm is also almost half of the $\lerr$ of \povmone, i.e., $9.6m$ vs $18.3m$.

\begin{figure}[h]
    \centering
    \includegraphics[width=0.38\textwidth]{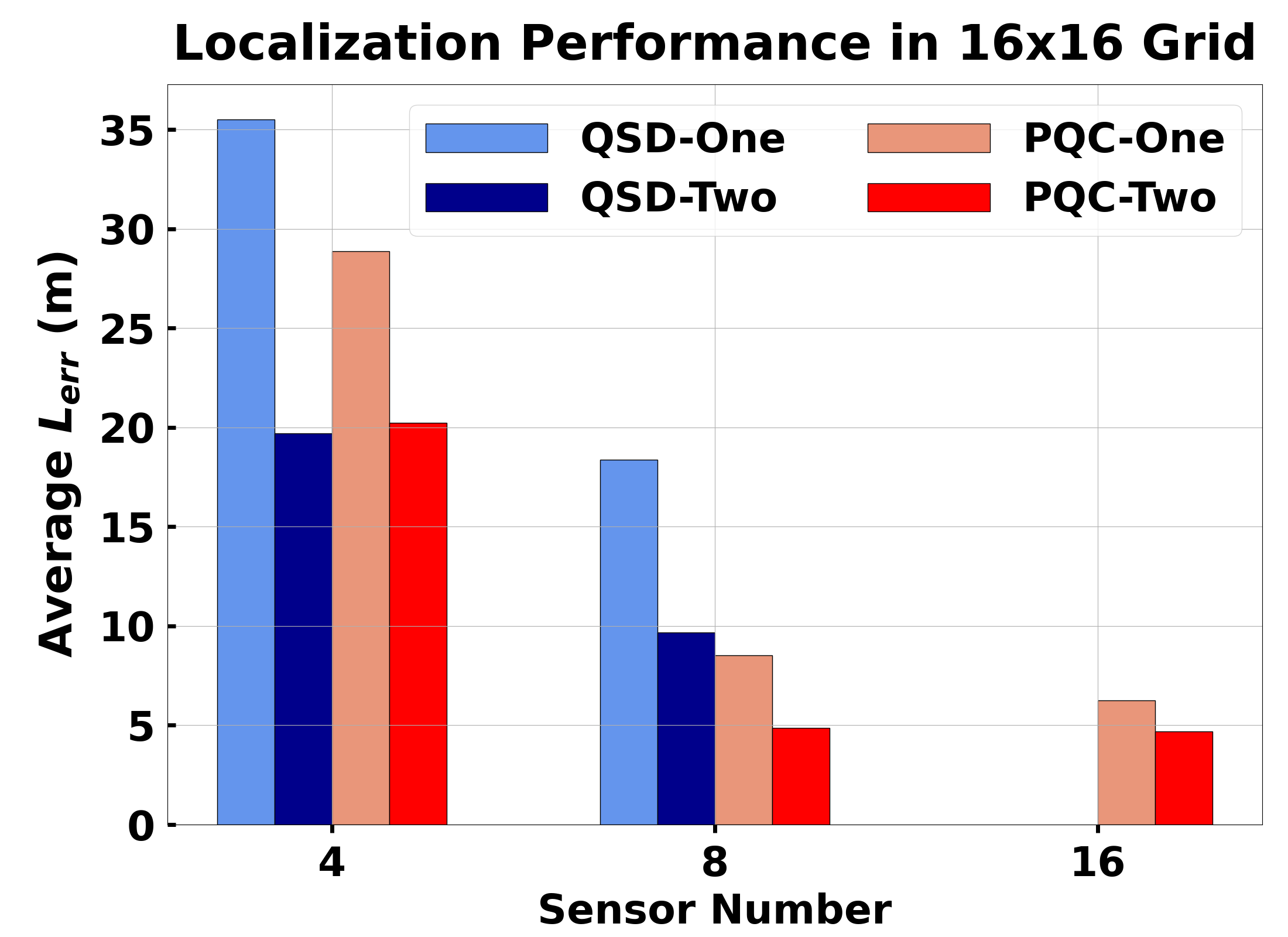}
    \caption{The performance of \povmone, \povm, \pqcone, \pqctwo for varying sensor number and a $16\times16$ grid.}
    \label{fig:continuous.varysen}
\end{figure}

\para{Varying Number of Sensors.} 
Fig.~\ref{fig:continuous.varysen} shows the average $\lerr$ in a $16\times16$ grid with a varying number of quantum sensors.
As expected, we observe that the $\lerr$ improves with an increasing number of quantum sensors,
for all four schemes.
For the \pqctwo scheme, we observe that the $\lerr$ improvement from 8 sensors to 16 sensors is minimal, i.e. $4.9 m$ vs $4.7 m$. 
This is because having 8 sensors in the first/coarse level seems sufficient to determine the block,
and then, in the second level, each block will always have 4 sensors associated with it (performance in the fine level is the same). 

\begin{figure}[h]
    \centering
    \includegraphics[width=0.38\textwidth]{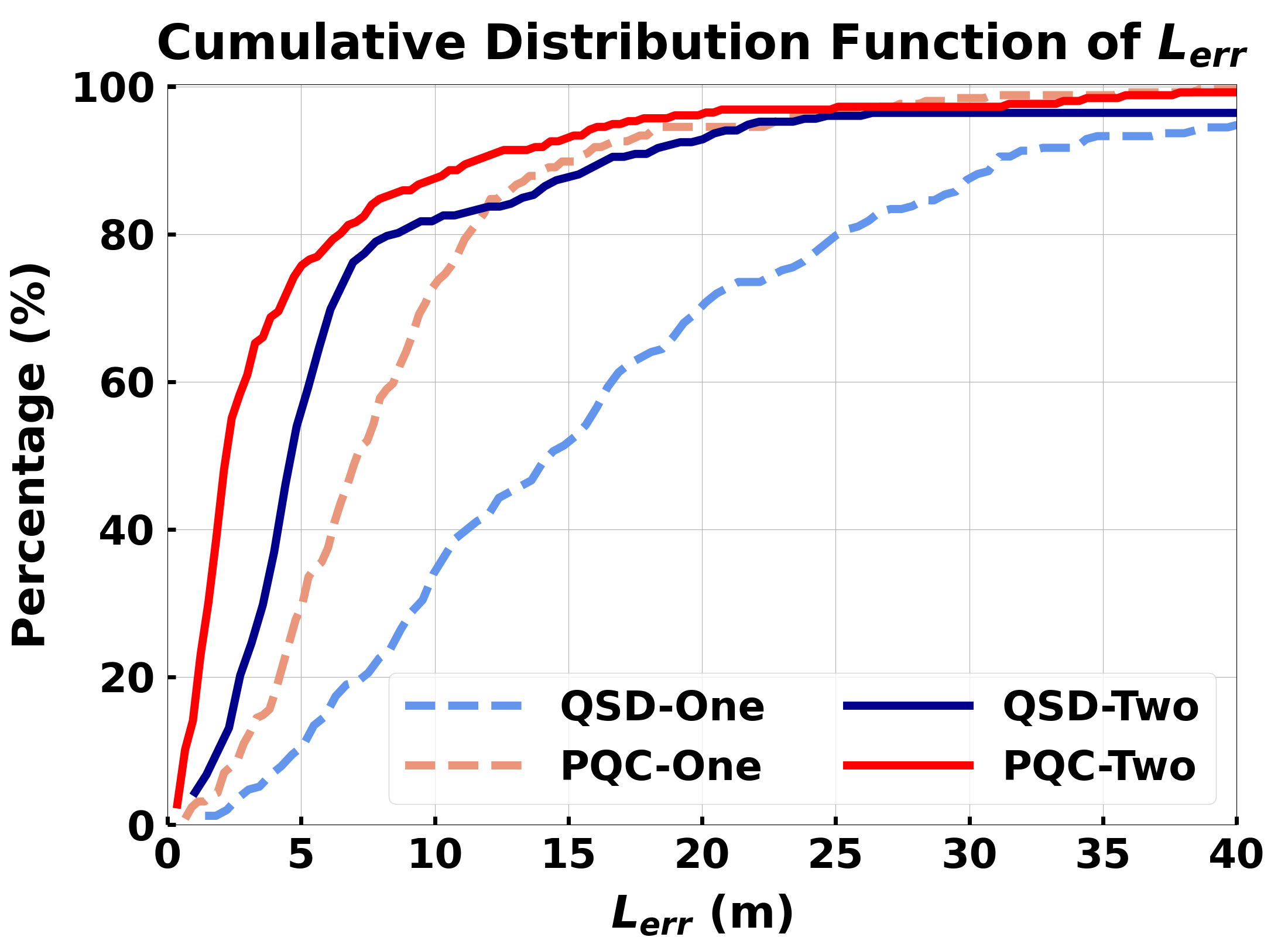}
    \caption{The cumulative probability of $\lerr$ of \povmone, \povm, \pqcone, \pqctwo for a $16\times16$ grid and 8 quantum sensors.}
    \label{fig:continuous.errorcdf}
\end{figure}

\para{CDF.}
Fig.~\ref{fig:continuous.errorcdf} shows the cumulative distribution function of $\lerr$ for four methods when the grid size is $16\times16$ and the number of sensors is 8.
This plot gives insight into the {\em distribution} of $\lerr$ over different TX location, 
compared with Fig.~\ref{fig:continuous.varysen} which shows only average $\lerr$ across many TX locations. 
Here, the distribution is over 256 TX locations---one random TX location per cell for 256 cells in the $16\times 16$ grid.
We observe, as expected, that  the two-level schemes are better than the one-level schemes, and the PQC-based methods are better than the QSD-based methods, except that \povm has a better CDF plot than
\pqcone up to the 83-th percentile. The above exception implies that \povm has higher number of locations with large $\lerr$ compared with \pqcone;
this is likely due to \povm incurring errors in determining the block at the first/coarse level, which can lead to large localization errors.





\para{Discrete Setting:} 
In the previous evaluation results, we have considered the practical {\em continuous} setting wherein the transmitter can be anywhere in the area.
To evaluate the true performance of our QSD-based methods, which are fundamentally classification strategies, we now evaluate the discrete setting
wherein the transmitter is located only at the center of a cell and the 
predicted output of a localization method is the cell number of the transmitter.
In this {\em discrete} setting, we evaluate the performance metric of {\em Classification Accuracy} $\lacc$ which is the percentage of times the method is correct in predicting the {\em cell} number. 
Also, in this discrete setting, the PQC-based methods use the Classification variant in the location predictor component, while the QSD-based methods remain the same.

\begin{figure}
    \centering
    \includegraphics[width=0.38\textwidth]{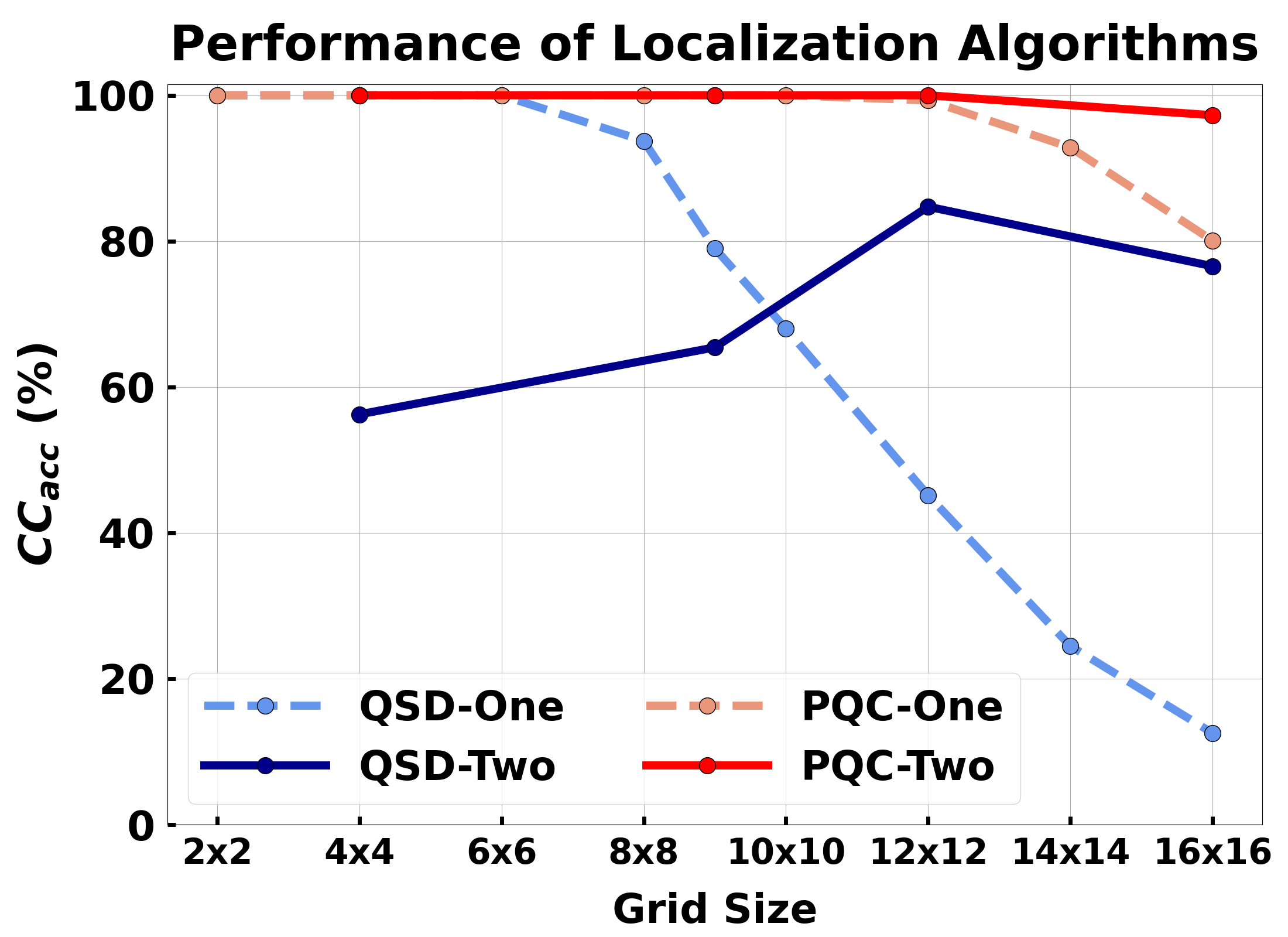}
    \caption{Performance of \povmone, \povm, \pqcone, \pqctwo for varying grid size and 8 sensors.}
    \label{fig:discrete.varygrid}
\end{figure}

Fig.~\ref{fig:discrete.varygrid} shows the performance of the four algorithms with varying grid sizes when the number of quantum sensors is eight. We observe similar trend for each algorithm as well as similar relative trends among the
algorithms as in the continuous setting. 
We make two important observations: 
\begin{enumerate}
    \item First, in the QSD-based methods, the \povm is a significant improvement over \povmone (from 13\% to 77\% for grid side $16\times16$), which shows the effectiveness of our two-level approach.
    \item Second, for the largest grid size of $16\times16$, the $\lacc$ for QSD-based \povm is reasonable at 77\% but is further 
improved impressively by \pqctwo at 97\%; this shows the effectiveness of our PQC-based methods. The 3\% error here in \pqctwo is mainly due to the errors in the first level of determining the block.
\end{enumerate}
Also, we see that for lower grid sides, the \povm surprisingly performs worse than \povmone; the 
reason for this is similar to the continuous case that determining the blocks at the first level becomes more erroneous when the grid size is small.

\begin{figure}[h]
    \centering
    \includegraphics[width=0.38\textwidth]{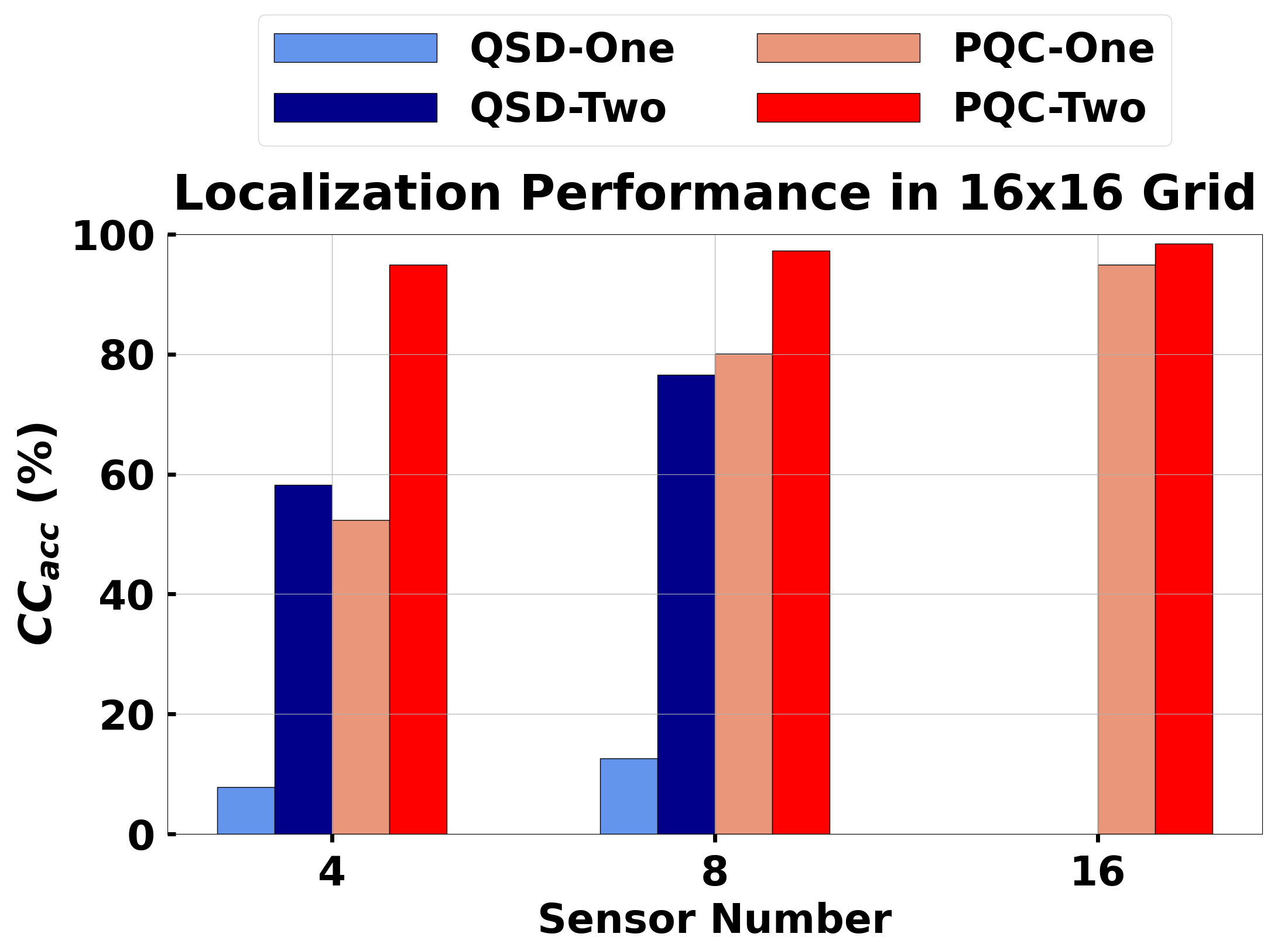}
    \caption{Performance of \povmone, \povm, \pqcone, \pqctwo for varying sensor number and a $16\times16$ grid.}
    \label{fig:discrete.varysen}
\end{figure}


Fig.\ref{fig:discrete.varysen} shows the $\lacc$ in a $16\times16$ grid for varying number of quantum sensors. As expected, we observe that the $\lacc$ improves with an increasing number of quantum sensors. More importantly, the $\lacc$ for \pqctwo is {\em near-perfect} 
at $99\%$ with 16 sensors; this shows the effectiveness of the two-level method as well
as of the well-trained parameterized hybrid circuits. As in the continuous-domain setting, we don't show the QSD-methods for 16 sensors, as it was infeasible to implement the QSD-based methods for a large number of sensors.



\section{\bf Conclusion and Future Work}
\label{sec:conclusion}

In this paper, we have developed effective schemes for an important quantum sensor
network application, viz., localization of a wireless transmitter. The work demonstrates
how a network of quantum sensors can collaborate to predict a parameter (here, location of an
event/transmitter) that is received/sensed differently at different sensor locations (e.g., depending on the distance from the event).
In particular, this work shows the promise of quantum sensor networks in localization of 
events in general---one of the most important applications of classical sensor networks.

Our work has significant opportunities for improvement. 
In particular, one can optimize the initial state of the sensors to further improve the localization performance (note that, here, we have only used a uniform superposition initial state).
In this context, we are also interested in determining whether
entangled initial states are helpful; recent 
works~\cite{swapping-tqe-22,predist-qce-22,ghz-qce-23} have shown that entangled states can be efficiently distributed over a quantum network. 
In addition, one can consider general multi-level approaches and restricted forms of measurement,
design parameterized quantum circuits with noise~\cite{quantumnas2022}, 
and develop techniques to distribute such circuits over a quantum (sensor) network as in~\cite{dqc-disc-q21, gdqc-qce-22, dqc-qsw-23}.
We plan to explore some of these directions in our future work.


\section*{Acknowledgment}
This work was supported by NSF awards FET-2106447 and CNS-2128187.

\bibliographystyle{IEEEtran}
\bibliography{bib}

\end{document}